# Fractional Quantum Anomalous Hall Effect in Multilayer Graphene

Zhengguang Lu[1]†, Tonghang Han[1]†, Yuxuan Yao[1]†, Aidan P. Reddy[1], Jixiang Yang[1], Junseok Seo[1], Kenji Watanabe[2], Takashi Taniguchi[3], Liang Fu[1], Long Ju[1]*

[1]Department of Physics, Massachusetts Institute of Technology, Cambridge, MA, USA.

[2]Research Center for Electronic and Optical Materials, National Institute for Materials Science, 1-1 Namiki, Tsukuba 305-0044, Japan

[3]Research Center for Materials Nanoarchitectonics, National Institute for Materials Science, 1-1 Namiki, Tsukuba 305-0044, Japan

*Corresponding author. Email: longju@mit.edu †These authors contributed equally to this work.

**The fractional quantum anomalous Hall effect (FQAHE), the analog of the fractional quantum Hall effect[1] at zero magnetic field, is predicted to exist in topological flat bands under spontaneous time-reversal-symmetry breaking[2–6]. The demonstration of FQAHE could lead to non-Abelian anyons which form the basis of topological quantum computation[7–9]. So far, FQAHE has been observed only in twisted MoTe$_2$ (t-MoTe$_2$) at moiré filling factor $v > 1/2$[10–13]. Graphene-based moiré superlattices are believed to host FQAHE with the potential advantage of superior material quality and higher electron mobility. Here we report the observation of integer and fractional QAH effects in a rhombohedral pentalayer graphene/hBN moiré superlattice. At zero magnetic field, we observed plateaus of quantized Hall resistance $R_{xy} = \frac{h}{ve^2}$ at filling factors $v$ = 1, 2/3, 3/5, 4/7, 4/9, 3/7 and 2/5 of the moiré superlattice respectively. These features are accompanied by clear dips in the longitudinal resistance $R_{xx}$. In addition, at zero magnetic field, $R_{xy}$ equals $\frac{2h}{e^2}$ at $v$ = 1/2 and varies linearly with the filling factor—similar to the composite Fermi liquid (CFL) in the half-filled lowest Landau level at high magnetic fields[14–17]. By tuning the gate displacement field $D$ and $v$, we observed phase transitions from CFL and FQAH states to other correlated electron states. Our graphene system provides an ideal platform for exploring charge fractionalization and (non-Abelian) anyonic braiding at zero magnetic field[7–9,18–20], especially considering a lateral junction between FQAHE and superconducting regions in the same device[21–23].**

The fractional quantum Hall effect observed in conventional two-dimensional electron gas (2DEGs) at the semiconductor interface is a classic example of intertwined electron correlation and topology effects in condensed matter physics[1]. It has been proposed that similar exotic states could exist at zero magnetic field, by engineering flat electronic bands with non-zero Chern numbers [2–6]. As a precursor of FQAHE, the integer quantum anomalous Hall effect (IQAHE) has been conceived[24] and realized in magnetic topologic insulators and the moiré superlattice made of two-dimensional materials[25–29]. Recently, FQAHE has been proposed in twisted transition metal dichalcogenides (TMD) moiré superlattices[30–34] and observed in twisted MoTe$_2$ at filling factors $v > 1/2$ [10–13]. Graphene-based moiré superlattices have also been proposed to host FQAH states[35–40], and could potentially show a plethora of fractional states due to the fewer defects in graphene lattice than in TMD lattices. Although fractional Chern insulators have been observed in graphene-based moiré superlattices at high magnetic fields[41,42], observation of FQAHE in any graphene system has not been reported to date.

The moiré superlattice formed between rhombohedral graphene (RG) and hexagonal boron nitride (hBN) has been demonstrated to be a remarkable platform for emergent quantum phenomena. In the highly tunable electronic bands in crystalline multilayer RG[43–46], correlated insulators, superconductivity, Chern insulators, orbital magnetism and multiferroicity have been demonstrated[43,46–53]. When placed on hBN to form a moiré superlattice, RG exhibits Mott insulators, tunable ferromagnetism and Chern insulators, as well as superconductivity[29,54–56]. Based on a tight-binding calculation, the band dispersion at zero gate-displacement field $D$ becomes flatter as the layer number increases from $N = 2$ to 5, while the valence band starts to suffer from increased trigonal warping with $N > 5$ [57]. At non-zero $Ds$, states near the band edge acquire larger Berry curvatures[57,58] that enable the formation of topological flat band when a moiré potential from hBN is introduced[38,59]. Therefore, engineering topological flat bands in gate-tuned multilayer RG/hBN is a promising approach for realizing FQAHE. From an experimental point of view, the RG/hBN moiré superlattice features three advantages over twisted TMD moiré superlattices: 1. the higher material quality of graphene over TMD; 2. better electrical contact to graphene than to TMD; 3. the twist-angle inhomogeneity induces less variations of the moiré period in a hetero-bilayer moiré than in a homo-bilayer moiré. While the Chern insulator state has been observed in trilayer RG/hBN at integer moiré fillings[29,49], FQAHE has not been found in this system so far.

Here we report observations of IQAHE and FQAHEs in a new graphene moiré system formed by pentalayer RG and hBN, as shown in Fig. 1a. While the general possibility of FQAHE in RG/hBN moiré superlattices has been suggested before[38], the pentalayer system we study has not received any concrete theoretical analysis or predictions. The superlattice period is ~11.5 nm and the twist angle is ~0.77°. We observed quantized Hall resistance $R_{xy} = \pm\frac{h}{e^2}$ at a moiré superlattice filling factor $v = 1$ and zero magnetic field. These states correspond to Chern numbers $C = \pm 1$. At fractional filling factors between 0 and 1, we observed 6 states with fractionally quantized Hall resistances $R_{xy} = \frac{h}{\nu e^2}$ at zero magnetic field. We observed $R_{xy} = \frac{2h}{e^2}$ at $\nu = 1/2$ and a linear dependence of $R_{xy}$ on $\nu$. By tuning $D$, we observed phase transitions from composite Fermi liquid (CFL) to valley-polarized Fermi liquid and a correlated insulator. We have measured two devices with similar moiré superlattice periods and they both show IQAHE and FQAHE. The data presented in the main text is based on Device 1, while the data from Device 2 is included in Extended Data Figures 8&9.

**Phase Diagram of the Moiré Superlattice**

Figure 1b&c show the longitudinal resistance $R_{xx}$ and transverse resistance $R_{xy}$ as functions of charge density $n_e$ (filling factor $\nu$) and $D$ measured at temperature $T = 10$ mK. $R_{xx}$ and $R_{xy}$ have been symmetrized and anti-symmetrized using data collected at $B = \pm 0.1$ T (see Methods and Extended Data Figure 4). While most regions on the maps show small $R_{xx}$ and $R_{xy}$, large values of resistances emerge in a tilted stripe region. At small $\nu$ up to 1/2, large $R_{xx}$ emerges, while $R_{xy}$ shows large fluctuations around zero. At $\nu = 2/5$ to 1, large $R_{xy}$ emerges and gradually changes with $\nu$. At $\nu = 1$, 2/3, 3/5, and 2/5, $R_{xx}$ shows local minimums while $R_{xy}$ shows plateaus as a function of $\nu$. As we show in the following sections, these states feature quantized $R_{xy}$ as expected for IQAHE and FQAHEs. We note that the charge density corresponding to $\nu = 1$ in our system is about 5 times smaller than that in t-MoTe$_2$[10–13], due to the larger moiré superlattice period.

The insulating state with large $R_{xx}$ we observed at $\nu \leq 1/2$ occupies a large continuous range of filling factor, which is distinct from correlated insulators at integer filling factors or generalized Wigner crystals at discrete fractional filling factors of moiré superlattices[60,61]. It is possible, for example, that electron crystallization happens within the flat moiré conduction band and leads to a Wigner crystal state at zero magnetic field[62]. At $D < 0$, we observed correlated insulating states at integer filling factors $\nu = 2$, 3 and 4 (see Extended Data Figure 1). The existence of topological states and correlated insulating states at opposite $D$s are consistent with previous experiments on RG/hBN moiré superlattices[29,49,54]. In this work, we focus on the regime of large positive displacement fields.

In the tilted stripe region, the large anomalous Hall signals indicates spontaneous valley polarization that breaks the time-reversal symmetry. As the $D$ required to observe the anomalous Hall signal is quite high, at finite charge density, the re-distribution of charges is expected to partially screen the externally applied $D$. This picture may explain the slope of the striped region, in which a larger $D$ is needed to maintain a narrow bandwidth as $\nu$ increases. The observation of IQAHE and FQAHEs at $\nu <= 1$ further suggests the presence of a topological flat band.

Fig 1d shows a moiré band structure calculated from a tight-binding model for pentalayer RG[59], with an added phenomenological superlattice potential to account for moiré effects from the nearly aligned hBN layer. For a top-to-bottom interlayer potential difference $\Delta$ = 75 meV, this model produces a $|C|=1$ lowest conduction moiré band that is extremely narrow (bandwidth < 5 meV) and isolated from other bands with a global band gap. In contrast, the lowest valence moiré band is significantly broader (bandwidth > 45 meV), topologically trivial ($C = 0$), and energetically overlapping with adjacent moiré bands.

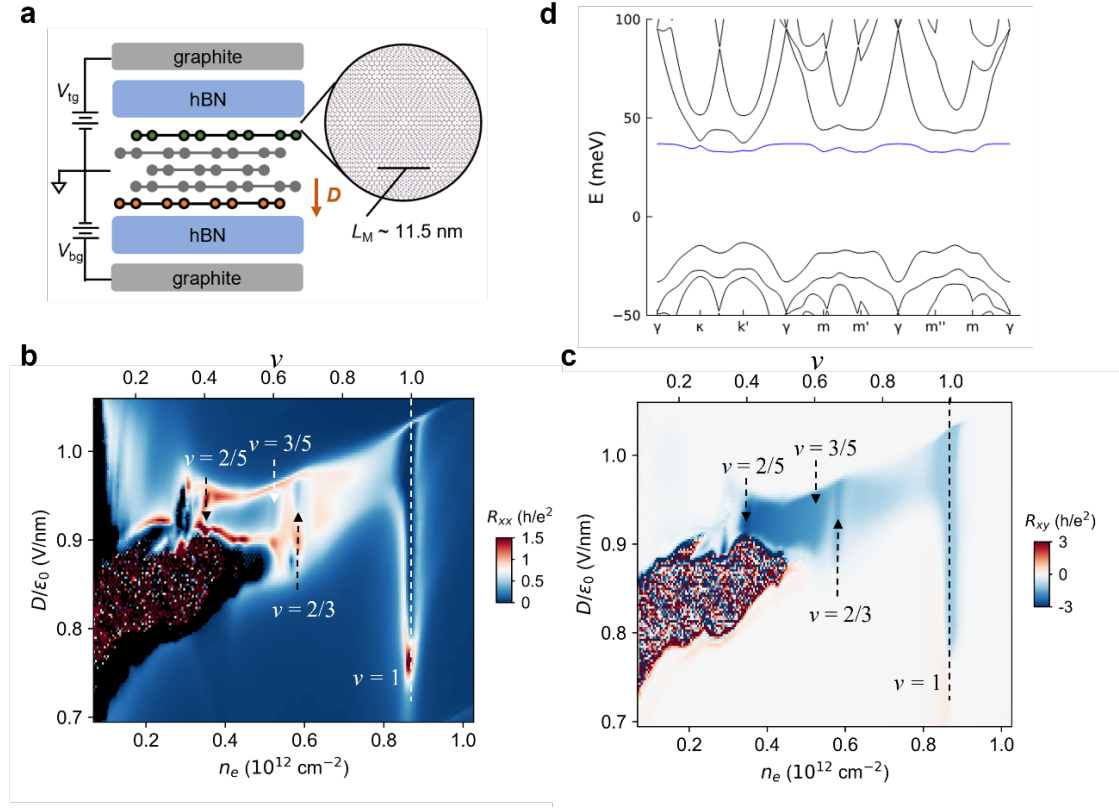

*Fig. 1. Device configuration, topological flat band and phase diagram of the rhombohedral pentalayer graphene/hBN moiré superlattice. a. Schematic of the device configuration, showing a moiré superlattice between the top layer of graphene and the top hBN, with a moiré period of 11.5 nm. b-c. Phase diagrams of the device revealed by symmetrized $R_{xx}$ and anti-symmetrized $R_{xy}$ at B = ± 0.1 T as functions of $n_e$ (v) and D. The temperature at the mixing chamber of dilution refrigerator is 10 mK. Large anomalous Hall signals emerge in a tilted stripe region centered at $D/\varepsilon_0 \sim 0.93$ V/nm. Clear dips of $R_{xx}$ can be seen at filling factors of the moiré superlattice v = 1, 2/3, 3/5 and 2/5 (indicated by the dashed lines and arrows), where $R_{xy}$ shows plateaus of values. d. Calculated band structure of our moiré superlattice with interlayer potential $\Delta$ = 75 meV, showing a flat $|C|=1$ moiré conduction band and a dispersive C=0 moiré valence band.*

**Integer Quantum Anomalous Hall Effect**

Figure 2 shows detailed characterizations of the anomalous Hall state at $v = 1$. At $D/\varepsilon_0 = 0.97$ V/nm, both $R_{xy}$ and $R_{xx}$ exhibit hysteretic behaviors under scanned magnetic field, as shown in Fig. 2a&b. At $T = 0.1$ K, $R_{xy}$ is quantized at $\frac{h}{e^2}$ at zero magnetic field, while $R_{xx}$ is smaller than 5 Ω. $R_{xy}$ remains quantized up to at least 1.6 K and $R_{xx}$ remains small in the same temperature range. Figure 2c&d show the $B$-dependence of $R_{xy}$ and $R_{xx}$ in the vicinity of $v = 1$ at $D/\varepsilon_0 = 0.97$ V/nm. The anomalous Hall state persists to $B = 0$ T and exhibits a wide plateau in both $R_{xy}$ and $R_{xx}$. The dispersion of this state agrees well with a Chern number $C = \pm 1$ state (indicated by the dashed line) according to the Streda's formula. Starting from ~0.6 T, more dips in $R_{xx}$ emerge and their slopes with $B$ agree with $C = 2$ and 3 states (indicated by additional dashed lines). Figure 2e shows the $n_e$ ($v$) dependence of $R_{xy}$ and $R_{xx}$ and features a plateau of width $\Delta n_e \sim 3*10^{10}$ cm$^{-2}$. The quantized $R_{xy}$ and small $R_{xx}$ values of the state at $v = 1$ exist in a wide range of $D$, as shown in Fig. 2f. At both higher and lower $D$s, the device shows small $R_{xx}$ and $R_{xy}$, except for a peak of $R_{xx}$ during both transitions.

These observations indicate the IQAHE with $C = \pm 1$ at filling factor $v = 1$. The features corresponding to $C = 2$ and 3 are due to integer quantum Hall effects that emerge at a low magnetic field, which demonstrates the high electron mobility of our device. The incremental change $\Delta C = 1$ between these three features indicates that the isospin degeneracy is completely lifted at $v = 1$, which corresponds to one electron per moiré unit cell. The width of the IQAHE plateau corresponds to a ~10 times smaller charge density than that in t-MoTe$_2$[11,13].

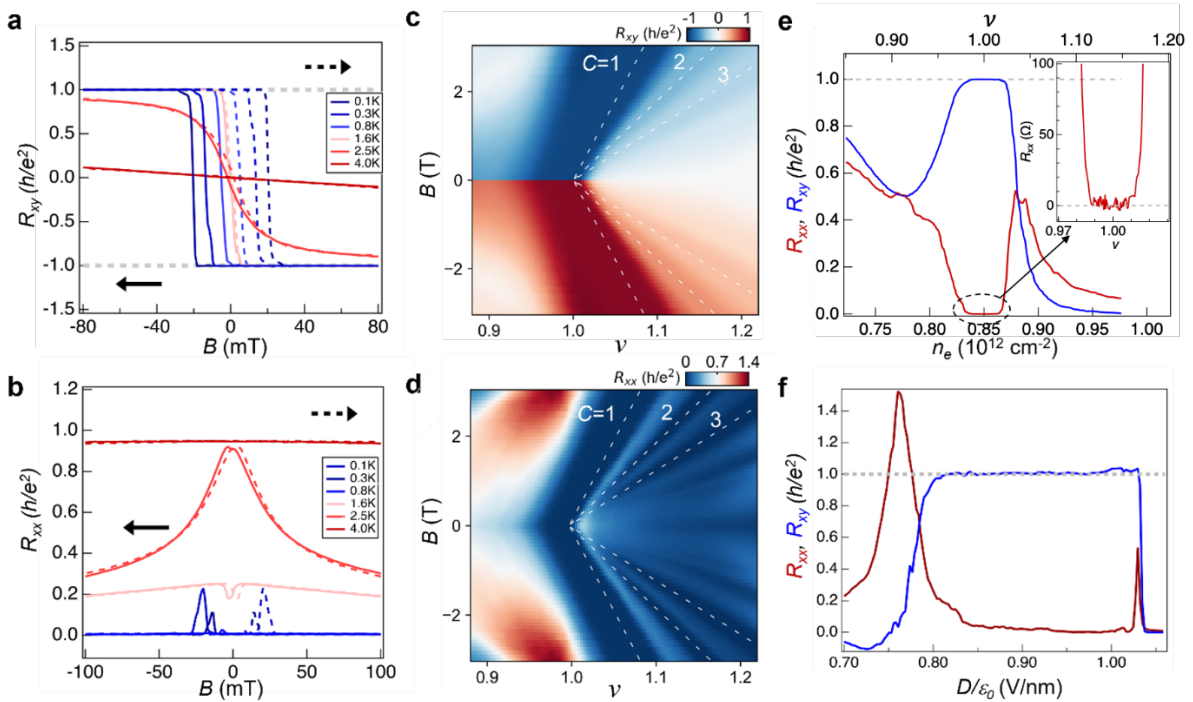

*Fig. 2. Integer quantum anomalous Hall effect. a & b.* Magnetic hysteresis scans of $R_{xy}$ and $R_{xx}$ at $v = 1$ and $D/\varepsilon_0 = 0.97$ V/nm and $T = 0.1 - 4$ K. Solid (dashed) lines correspond to scanning B from positive (negative) values to negative (positive) values. At 0.1 K, $R_{xy}$ is quantized at $\pm\frac{h}{e^2}$ which corresponds to a Chern number $C = \pm 1$, while $R_{xx}$ shows a value $< 5$ Ω at $B = 0$ mT. *c-d.* Landau-fan diagrams of $R_{xx}$ and $R_{xy}$ at $D/\varepsilon_0 = 0.97$ V/nm. The IQAH state can be seen as a wide plateau in both maps, which disperses with magnetic field with a slope that agrees well with the dashed lines (corresponding to $C = \pm 1$, as determined by the Streda's formula). At above ~0.6 T, features associated with integer quantum Hall states start to appear, corresponding to $C = \pm 2$ and 3 correspondingly as indicated by additional dashed lines. *e.* Symmetrized $R_{xx}$ and anti-symmetrized $R_{xy}$ (we present the positive values for convenience) as functions of $v$ at $T = 10$ mK and $D/\varepsilon_0 = 0.97$ V/nm, featuring quantized $R_{xy}$ in a plateau of width $\sim 3*10^{10}$ cm$^{-2}$. Inset: zoomed-in plot to reveal the plateau of $R_{xx} < 5$ Ω. *f.* $R_{xx}$ and $R_{xy}$ as a function of D at $v = 1$, featuring a wide plateau at ~0.8-1.03 V/nm. Moving to both higher and lower Ds, the device transitions to metallic states with small $R_{xx}$ and $R_{xy}$. A peak in $R_{xx}$ appears during both transitions.

**Fractional Quantum Anomalous Hall Effects**

As shown by Fig. 1b&c, large anomalous Hall response is found in a wide range between $v = 2/5$ and 1. Figure 3a&b show finer maps of $R_{xx}$ and $R_{xy}$ in this range, where additional vertical line features can be seen. To better visualize the states corresponding to these lines, we present in Fig. 3c line-cuts along the dashed lines in Fig. 3a&b. We can observe plateaus of $R_{xy}$ at $v = 2/5, 3/7, 4/9, 4/7, 3/5$ and $2/3$ with the value of $R_{xy}$ quantized at $\frac{h}{ve^2}$. At the same time, $R_{xx}$ shows clear dips at these filling factors, similar to the observations of fractional quantum Hall states in 2DEGs at high magnetic fields[1,14]. We note that the $v = 2/5$ state is right next to the boundary of the anomalous Hall region, which shows a peak in $R_{xx}$. Nevertheless, the clear plateau of $R_{xy}$ and dip in $R_{xx}$ are observed at $v = 2/5$. Figure 3d-f & h-j show the magnetic hysteresis scans at fractional filling factors corresponding to the states identified in Fig. 3c. For all these states, $R_{xy}$ shows quantized values of $\frac{h}{ve^2}$ while $R_{xx}$ is much smaller than $R_{xy}$. Lastly, Fig. 3g&k shows the Landau-fan diagram of $R_{xx}$ in the range of $v < 1/2$ and $v > 1/2$, respectively. The dips at fractional filling factors evolve into tilted lines whose slopes agree well with the dashed lines, which are calculated based on the Streda's formula $\frac{\partial n_e}{\partial B} = \frac{ve}{h}$ for the corresponding filling factors. As a function of D, all FQAH states develops a plateau of $R_{xy}$ at the corresponding quantized values (see Extended Data Figure 2) and dips of $R_{xx}$. The center of the $R_{xy}$ plateau and the dips in $R_{xx}$ shift to higher D as the filling factor increases, which agrees with the tilted stripe shape of the anomalous Hall region as shown in Fig. 1b&c.

The observations of quantized $R_{xy}$ and the corresponding dips in $R_{xx}$ at fractional filling factors, together with the hysteresis enclosing zero magnetic field indicate FQAH states in our graphene-based moiré superlattice. These states resemble the Jain sequence of fractional quantum Hall states[14–17], but at zero magnetic field. Compared with t-MoTe$_2$ where FQAHEs are only observed at $v > 1/2$[10–13], the fractional states we observed reside at both sides of the half-filling. This is likely due to the better electrical contact in graphene devices than that has been achieved in semiconductor devices at low charge densities. The narrowest plateau width of FQAH states we have observed is ~$10^{10}$ cm$^{-2}$, which is about 10 times narrower than the 3/5 state observed in t-MoTe$_2$[11].

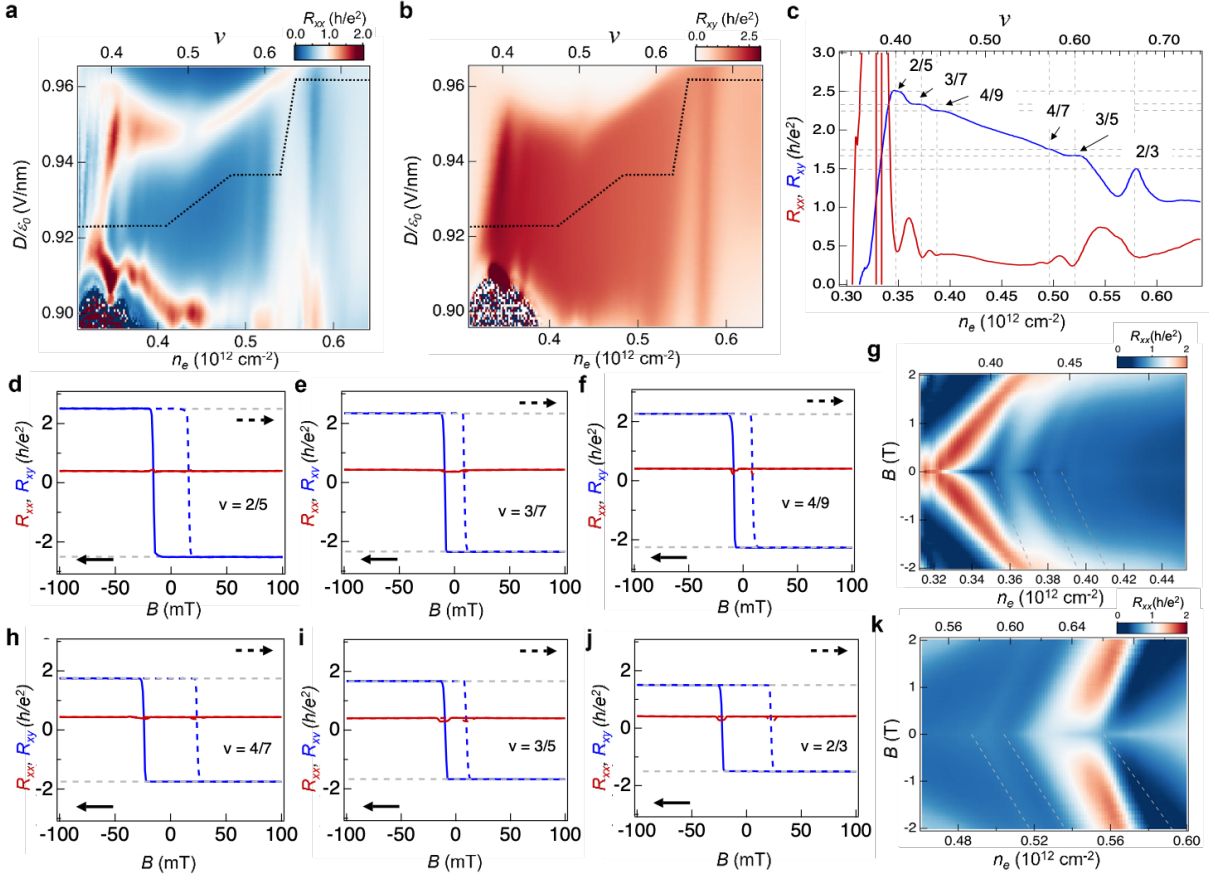

**Fig. 3. Fractional quantum anomalous Hall effects. a & b.** *Zoomed-in diagrams of symmetrized $R_{xx}$ and anti-symmetrized $R_{xy}$ (we present the positive values for convenience) at $B = \pm 0.1$ T as functions of $v$ ($n_e$) and D. Fine features that could not be identified in Fig. 1d&e can be seen in the vicinity of $v = 1/2$, especially in the $R_{xx}$ diagram. Data is collected using a constant voltage bias measurement.* **c.** *$R_{xx}$ and $R_{xy}$ along the dashed lines in **a** & **b**, taken with a constant current measurement. Clear plateaus of $R_{xy}$ at $\frac{5h}{2e^2}$, $\frac{7h}{3e^2}$, $\frac{9h}{4e^2}$, $\frac{7h}{4e^2}$, $\frac{5h}{3e^2}$ and $\frac{3h}{2e^2}$ emerge at $v = 2/5, 3/7, 4/9, 4/7, 3/5$ and $2/3$, as indicated by the dashed lines and*

*arrows. $R_{xx}$ shows clear dips at the corresponding filling factors.* **d-f & h-j.** *Magnetic hysteresis scans of $R_{xy}$ and $R_{xx}$ at v = 2/5, 3/7, 4/9, 4/7, 3/5 and 2/3, showing quantized values of $R_{xy} = \frac{h}{ve^2}$ and much smaller $R_{xx}$.* **g&k.** *Landau-fan diagram of $R_{xx}$ at $D/\varepsilon_0$ = 0.92 V/nm. The FQAH states can be seen as tilted line features, the slopes of which agree well with the dashed lines. The slopes of dashed lines correspond to C = 2/5, 3/7 and 4/9 in* **g**, *and 4/7, 3/5 and 2/3 in* **k** *using the Streda's formula.*

## AHE and Phase Transitions at Hall-filling

In addition to FQAHE states that have plateaus in $R_{xy}$ and dips in $R_{xx}$, we observed a continuously changing anomalous Hall resistance in a wide range of filling factors from 4/9 to 4/7. Especially at around v = 1/2, $R_{xy}$ varies roughly linearly with $n_e$ and v while $R_{xx}$ does not show any clear dips, as shown in Fig. 4a. At v = 1/2, $R_{xy}$ equals $\frac{2h}{e^2}$. Fig. 4b shows the hysteretic behaviors of $R_{xy}$ and $R_{xx}$ under a sweeping magnetic field. Figure 4c shows the $D$-dependence of $R_{xy}$ and $R_{xx}$ at fixed filling factor v = 1/2. The value of $R_{xy}$ spans in a plateau which is concurrent with small values of $R_{xx}$. At higher $D$, $R_{xy}$ decreases monotonically while $R_{xx}$ first increases and then decreases. At lower $D$, $R_{xx}$ shoots up and $R_{xy}$ shows large fluctuations in the range of 0.85 to 0.9 V/nm. At even lower $D$, $R_{xy}$ becomes almost zero while $R_{xx}$ remains at a few k$\Omega$. At higher $D$ than that of the $R_{xy} = \frac{2h}{e^2}$ plateau, $R_{xy}$ decreases monotonically while $R_{xx}$ first increases and then decreases. As shown in Fig. 4d, the state at $D/\varepsilon_0$ = 0.97 V/nm features small values of $R_{xy}$ and $R_{xx}$, but a decent value of the Hall angle $\theta_H$ ~9.5° corresponding to $\tan\theta_H = \frac{R_{xy}}{R_{xx}}$ ~0.17.

The absence of a dip in $R_{xx}$ at v = 1/2 and the linear dependence of $R_{xy}$ with v in the neighborhood are distinct from FQAH states we described in the previous section and suggest the absence of a charge gap[11,63,64]. These latter properties are reminiscent of the CFL in the half-filled lowest Landau level of 2DEGs at high magnetic fields[14–17]. Starting from the zero-magnetic-field CFL state, our data suggests two distinct types of phase transitions driven by $D$. At the higher $D$ side of the CFL state, the system is in a valley-polarized metallic state[49,51–53], due to the non-zero anomalous Hall resistance and small $R_{xx}$. The persistence of valley polarization and the peak of $R_{xx}$ at intermediate $D$s suggest that it is a continuous phase transition from CFL to Fermi liquid (FL). This new type of phase transition has been recently proposed by theory in FQAHE systems but was not observed in t-MoTe$_2$[65,66]. At the lower $D$ side, we have a phase transition from CFL to our observed correlated insulator at v ≤ 1/2. Our observations call for further experiments to explore both types of phase transition, which are beyond the scope of this work.

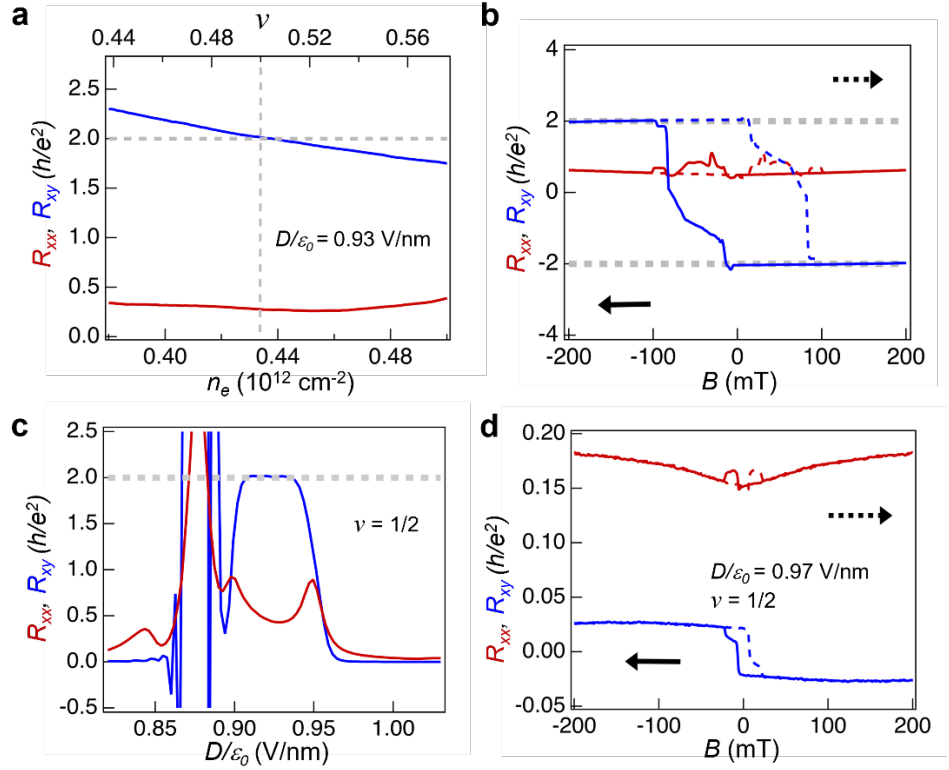

*Fig. 4. Anomalous Hall effect and phase transitions at half-filling. a. Symmetrized $R_{xx}$ and anti-symmetrized $R_{xy}$ (we present the positive values for convenience) at $B = \pm 0.1$ T and $D/\varepsilon_0 = 0.93$ V/nm in the neighborhood of half-filling. $R_{xy}$ shows a value of $\frac{2h}{e^2}$ at $v = 1/2$ and varies roughly linearly with the change of filling factor, while no dip in $R_{xx}$ is observed. These observations resemble the signatures of composite Fermi liquid (CFL) in 2DEGs at high magnetic fields. b. Magnetic hysteresis scans of $R_{xy}$ and $R_{xx}$ at $v = 1/2$, showing $R_{xy}$ plateaus at $\pm\frac{2h}{e^2}$ and much smaller $R_{xx}$. c. $R_{xy}$ and $R_{xx}$ at $v = 1/2$ as functions of D, showing the same $R_{xy}$ value in a plateau spanning from 0.9 to 0.94 V/nm. At higher D, both $R_{xy}$ and $R_{xx}$ decrease to close to zero. At lower D, $R_{xx}$ shoots up while $R_{xy}$ shows large fluctuations at around zero. d. Magnetic hysteresis scans of $R_{xy}$ and $R_{xx}$ at $v = 1/2$ and $D/\varepsilon_0 = 0.97$ V/nm, showing anomalous Hall signals and a Hall angle $\theta_H \sim 9.5°$, corresponding to $\tan\theta_H = \frac{R_{xy}}{R_{xx}} \sim 0.17$. This indicates a phase transition from CFL to valley-polarized metal at the higher D side. At the lower D side, the phase transition happens between CFL and a correlated insulating state.*

## Conclusion and Outlook

We observed IQHAE at $v = 1$ and FQAHEs at both sides of half-filling of the first moiré conduction band. Beyond the specific moiré superlattice demonstrated here, our results indicate the great potential of similar

RG/hBN systems with varied layer number, gate displacement field and twist angle for FQAHE studies—an opportunity that has been largely overlooked by theory and experiment so far. Given the high material quality, additional opportunities of researching novel quantum phase transitions, electron crystals at zero magnetic field, and behaviors of CFL in the moiré potential are within the reach of experiments[63,64]. The possibility of high-Chern-number flat bands in RG[51] also points to possibly more exotic FQAH states with non-Abelian anyons for topological quantum computation[7–9]. Furthermore, the co-existence of FQAHE and superconductivity[50,56] in graphene systems facilitates the realization of synthetic non-Abelian anyonic braiding by using a lateral junction within the same device[21–23].

**Data availability** The data shown in the main figures and other data that support the findings of this study are available from the corresponding authors upon reasonable request.

**Acknowledgments**

We acknowledge helpful discussions with X.G. Wen, T. Senthil, P. Lee, F. Wang and R. Ashoori. We thank D. Laroche for assistance with early investigation of a related sample. L.J. acknowledges support from a Sloan Fellowship. Work by T.H., J.Y. and J.S. was supported by NSF grant no. DMR- 2225925. The device fabrication of this work was supported by the STC Center for Integrated Quantum Materials, NSF grant no. DMR-1231319 and was carried out at the Harvard Center for Nanoscale Systems and MIT.Nano. Part of the device fabrication was supported by USD(R&E) under contract no. FA8702-15-D-0001. K.W. and T.T. acknowledge support from the JSPS KAKENHI (Grant Numbers 20H00354, 21H05233 and 23H02052) and World Premier International Research Center Initiative (WPI), MEXT, Japan. L.F. was supported by the STC Center for Integrated Quantum Materials (CIQM) under NSF award no. DMR-1231319.

**Author Contributions**

L.J. supervised the project. Z.L. and T.H. performed the DC magneto-transport measurement. T.H. and Y.Y. fabricated the devices. J.Y., J.S., Z.L. and T.H. helped with installing and testing the dilution refrigerator. A.R. and L.F. performed the calculations. K.W. and T.T. grew hBN crystals. All authors discussed the results and wrote the paper.